\documentclass[twocolumn,english,superscriptaddress,citeautoscript,showpacs,preprintnumbers,amsmath,amssymb,prb,floatfix,footinbib]{revtex4}
\usepackage[utf8x]{inputenc}
\usepackage[scaled=2]{helvet}
\usepackage[T1]{fontenc}
\usepackage[utf8]{luainputenc}
\usepackage{multirow}
\usepackage{textcomp}
\usepackage{amsmath}
\usepackage{amssymb}
\usepackage{graphicx}
\usepackage{subscript}
\usepackage{epstopdf}
\makeatletter

\newcommand{\lyxmathsym}[1]{\ifmmode\begingroup\def\b@ld{bold}
  \text{\ifx\math@version\b@ld\bfseries\fi#1}\endgroup\else#1\fi}

\providecommand{\tabularnewline}{\\}

\@ifundefined{textcolor}{}
{%
 \definecolor{BLACK}{gray}{0}
 \definecolor{WHITE}{gray}{1}
 \definecolor{RED}{rgb}{1,0,0}
 \definecolor{GREEN}{rgb}{0,1,0}
 \definecolor{BLUE}{rgb}{0,0,1}
 \definecolor{CYAN}{cmyk}{1,0,0,0}
 \definecolor{MAGENTA}{cmyk}{0,1,0,0}
 \definecolor{YELLOW}{cmyk}{0,0,1,0}
}

\usepackage[scaled=2]{helvet}\usepackage{amstext}

\makeatletter
\@ifundefined{textcolor}{}{\definecolor{BLACK}{gray}{0}\definecolor{WHITE}{gray}{1}\definecolor{RED}{rgb}{1,0,0}\definecolor{GREEN}{rgb}{0,1,0}\definecolor{BLUE}{rgb}{0,0,1}\definecolor{CYAN}{cmyk}{1,0,0,0}\definecolor{MAGENTA}{cmyk}{0,1,0,0}\definecolor{YELLOW}{cmyk}{0,0,1,0}}

\usepackage{dcolumn}
\usepackage{bm}
\usepackage{times}
\usepackage{hyperref}
\hypersetup{colorlinks=true,linkcolor=red,citecolor=blue}
\makeatother

\usepackage{babel}

\makeatother

\usepackage{babel}
\begin{document}
\title{Spin reorientation of the Fe moments in Eu$_{0.5}$Ca$_{0.5}$Fe$_{2}$As$_{2}$: Evidence for a strong interplay of Eu and Fe magnetism}

\author{W. T. Jin}
\email{jwt2006@gmail.com}
\affiliation{Jülich Centre for Neutron Science JCNS at Heinz Maier-Leibnitz Zentrum (MLZ), Forschungszentrum Jülich GmbH, Lichtenbergstraße 1, D-85747 Garching, Germany}

\author{M. Meven}
\affiliation{RWTH Aachen University, Institut für Kristallographie, D-52056 Aachen, Germany}
\affiliation{Jülich Centre for Neutron Science JCNS at Heinz Maier-Leibnitz Zentrum (MLZ), Forschungszentrum Jülich GmbH, Lichtenbergstraße 1, D-85747 Garching, Germany}

\author{A. P. Sazonov}
\affiliation{RWTH Aachen University, Institut für Kristallographie, D-52056 Aachen, Germany}
\affiliation{Jülich Centre for Neutron Science JCNS at Heinz Maier-Leibnitz Zentrum (MLZ), Forschungszentrum Jülich GmbH, Lichtenbergstraße 1, D-85747 Garching, Germany}

\author{S. Demirdis}
\affiliation{Jülich Centre for Neutron Science JCNS at Heinz Maier-Leibnitz Zentrum (MLZ), Forschungszentrum Jülich GmbH, Lichtenbergstraße 1, D-85747 Garching, Germany}

\author{Y. Su}
\affiliation{Jülich Centre for Neutron Science JCNS at Heinz Maier-Leibnitz Zentrum (MLZ), Forschungszentrum Jülich GmbH, Lichtenbergstraße 1, D-85747 Garching, Germany}

\author{Y. Xiao}
\affiliation{School of Advanced Materials, Peking University Shenzhen Graduate School, Shenzhen 518055, China}

\author{Z. Bukowski}
\affiliation{Institute of Low Temperature and Structure Research, Polish Academy of Sciences, 50-422 Wroclaw, Poland}

\author{S. Nandi}
\affiliation{Department of Physics, Indian Institute of Technology, Kanpur 208016, India}
\affiliation{Jülich Centre for Neutron Science JCNS and Peter Grünberg Institut PGI, JARA-FIT, Forschungszentrum Jülich GmbH, D-52425 Jülich, Germany}

\author{Th. Brückel}
\affiliation{Jülich Centre for Neutron Science JCNS and Peter Grünberg Institut PGI, JARA-FIT, Forschungszentrum Jülich GmbH, D-52425 Jülich, Germany}
\affiliation{Jülich Centre for Neutron Science JCNS at Heinz Maier-Leibnitz Zentrum (MLZ), Forschungszentrum Jülich GmbH, Lichtenbergstraße 1, D-85747 Garching, Germany}

\begin{abstract}
Using complementary polarized and unpolarized single-crystal neutron diffraction, we have investigated the temperature-dependent magnetic structures of Eu$_{0.5}$Ca$_{0.5}$Fe$_{2}$As$_{2}$. Upon 50 \% dilution of the Eu sites with isovalent Ca$^{2+}$, the Eu sublattice is found to be still long-range ordered below $\mathit{T_{Eu}}$ = 10 K, in the A-typed antiferromagnetic (AFM) structure. The moment size of Eu$^{2+}$ spins is estimated to be as large as 6.74(4) $\mu_{B}$ at 2.5 K. The Fe sublattice undergoes a spin-density-wave transition at $\mathit{T_{SDW}}$ = 192(2) K and displays an in-plane AFM structure above $\mathit{T_{Eu}}$. However, at 2.5 K, the Fe$^{2+}$ moments are found to be ordered in a canted AFM structure with a canting angle of 14(4)° out of the $\mathit{ab}$ plane. The spin reorientation of Fe below the AFM ordering temperature of Eu provides a direct evidence of a strong interplay between the two magnetic sublattices in Eu$_{0.5}$Ca$_{0.5}$Fe$_{2}$As$_{2}$.
\end{abstract}

\maketitle

\section{Introduction}

The discovery of superconductivity (SC) with the critical temperature $T_{c}$ = 26 K in fluorine-doped LaFeAsO in 2008 has opened up an ``iron age of superconductivity''.\cite{Kamihara_08} Shortly after that, $T_{c}$ above 50K was achieved in $\mathit{R}$FeAsO$_{1-x}$F$_{x}$ (``1111'' system, $\mathit{R}$ = Ce, Sm, Pr, Nd, and Gd) with $\mathit{R}$ being a rare-earth element.\cite{ChenXH_08,ChenGF_08,RenZA_08,WangC_08} SC with $T_{c}$ up to 38 K was also realized by various chemical substitutions in the ternary ``122'' compounds $\mathbf{\mathit{A}}$Fe$_{2}$As$_{2}$ with $\mathit{A}$ being an alkaline-earth-metal element (Ca, Ba, Sr) or the rare-earth element Eu.\cite{Rotter_08,Sefat_08,Zapf_17} It was well confirmed that the SC in the iron pnictides emerges upon the suppression of the static long-range spin-density-wave (SDW) order of Fe by means of chemical doping or applying external pressure.\cite{Johnston_10,Dai_15} Although SC is compatible with the localized moments of the rare-earth ions in either ``1111'' or ``122'' system,\cite{Ryan_09,Ren_09,Jeevan_11,Guguchia_13} how the magnetism of Fe and rare-earth element interact with each other is still not well elucidated.

In-depth experimental studies performed on quaterary ``1111'' $\mathit{R}$FeAsO system have provided compelling evidences that there is a strong coupling of Fe and $\mathit{R}$ magnetism for $\mathit{R}$ = Ce, Sm, Pr, and Nd, respectively.\cite{Maeter_09,ZhangQ_13,Nandi_11,Stockert_12,Tian_10} However, for ternary EuFe$_{2}$As$_{2}$ compounds, it is quite controversial regarding the strength of the interplay of Fe and Eu magnetism.\cite{Xiao_09,Herrero-Martin_09,Jeevan_08_parent,Pogrebna_15,Ahmed_10,Guguchia_11,Blachowski_11,Jin_16} As a special member of the ``122'' system, EuFe$_{2}$As$_{2}$ has drawn tremendous attention, due to the strong spin-charge-lattice coupling and doping- or pressure-induced coexistence of SC and strong ferromagnetism.\cite{Xiao_10,Xiao_12,Jin_13,Nandi_14,Nandi_14_neutron,Jin_15,Jin_PhaseDiagram,Jin_Pressure} In a purely ionic picture, the $\mathit{S}$-state (orbital moment $\mathit{L}$ = 0) Eu$^{2+}$ rare-earth ion has a 4$\mathit{f}$$^{7}$ electronic configuration and a total electron spin $\mathit{S}$ = 7/2, corresponding to a theoretical total effective magnetic moment of $\mu_{eff}$ = $\mathit{g}$$\sqrt{S(S+1)}$ = 7.94 $\mathit{\mu_{B}}$ (with the Land\'{e} factor $\mathit{g}$ = 2).\cite{Marchand_78} The non-superconducting parent compound EuFe$_{2}$As$_{2}$ undergoes a structural phase transition from a tetragonal to an orthorhombic phase at 190 K, concomitant with a SDW ordering of the itinerant Fe moments. In addition, the localized Eu$^{2+}$ spins order below 19 K in the A-type antiferromagnetic (AFM) structure (ferromagnetic layers stacked antiferromagnetically along the $\mathit{c}$ axis).\cite{Jiang_09_NJP} 

According to previous neutron and non-resonant x-ray magnetic scattering experiments,\cite{Xiao_09,Herrero-Martin_09} the coupling between the Eu and Fe sublattices in EuFe$_{2}$As$_{2}$ was found to be negligible, which was further supported by density-functional electronic structure calculations.\cite{Jeevan_08_parent} Also, a direct optical pump-probe showed a slow response of the Eu$^{2+}$ spins to the optical excitation of the itinerant carriers on the FeAs layers, suggesting a weak coupling between the two sublattices.\cite{Pogrebna_15} In contrast, magnetic Compton scattering on EuFe$_{2}$(As$_{0.73}$P$_{0.27}$)$_{2}$ indicated that the magnetism of Fe gets enhanced when the Eu magnetic order sets in.\cite{Ahmed_10} In addition, nuclear magnetic resonance (NMR) and Mössbauer spectroscopy measurements revealed a strong coupling between the localized Eu$^{2+}$ moments and the conduction $\mathit{d}$ electrons on the FeAs layers in Co-doped EuFe$_{2}$As$_{2}$.\cite{Guguchia_11,Blachowski_11} Recently, by performing x-ray resonant magnetic scattering (XRMS) measurement on underdoped Eu(Fe$_{0.94}$Ir$_{0.06}$)$_{2}$As$_{2}$, we have observed the magnetic polarization of the Ir 5$\mathit{d}$ band induced by the AFM ordering of Eu, indicating a strong interplay between the two sublattices.\cite{Jin_16} Undoubtedly, detailed knowledge about the evolution of magnetic structures of both Eu and Fe with the temperature will be crucial for understanding these observations.

Isovalent substitution of Eu with Ca offers an ideal platform for studying the delicate interplay between the two magnetic sublattices. On the one hand, under ambient pressure, Ca doping into the Eu site does not perturb the SDW order in the FeAs layers visibly and never
leads to SC. On the other hand, dilution of the Eu sublattice with nonmagnetic Ca$^{2+}$ ions suppresses its AFM ordering temperature ($\mathit{T_{Eu}}$) gradually.\cite{Mitsuda_11,Tran_18,Harnagea_18} A recent $\mu$SR study on Eu$_{0.5}$Ca$_{0.5}$Fe$_{2}$As$_{2}$ suggests a long-range magnetically ordered Eu sublattice.\cite{Tran_18} However, it was proposed based on macroscopic measurements that substitution of 50 \% Eu ions might lead to a short-range ordered nature of Eu
magnetism.\cite{Jeevan_08} In order to determine the ground-state magnetic structure of Eu$_{0.5}$Ca$_{0.5}$Fe$_{2}$As$_{2}$ and check the interplay between two sublattices, we have performed the temperature-dependent polarized and unpolarized neutron diffraction
studies on the Eu$_{0.5}$Ca$_{0.5}$Fe$_{2}$As$_{2}$ single crystal. The Eu$^{2+}$ moments are found to be long-range ordered below $\mathit{T_{Eu}}$ = 10 K, in the A-typed AFM structure. A spin-reorientation of the Fe$^{2+}$ moments is clearly observed around $\mathit{T_{Eu}}$, providing a direct evidence of a strong coupling between the Fe and Eu magnetism.

\section{Experimental Details}

Single crystals of Eu$_{1-x}$Ca$_{x}$Fe$_{2}$As$_{2}$ ($\mathit{x}$ = 0.5 nominally) were grown using the Sn flux method.\cite{Tran_18} No incorporation of Sn into the crystals was evidenced according to the energy-dispersive x-ray spectroscopy (EDX) characterization. The concentration of Ca was determined to be 52(4) \% by nuclear structure refinement of the neutron diffraction data, as presented below. A 88 mg platelike single crystal with dimensions $\sim$ 4 \texttimes{} 3 \texttimes{} 0.6 mm$^{3}$ was selected for unpolarized and polarized neutron diffraction measurements, which were performed on the hot-neutron four-circle diffractometer HEiDi and and diffuse scattering cold-neutron spectrometer DNS, respectively, at Heinz Maier-Leibnitz Zentrum (MLZ), Garching (Germany).\cite{Meven_15,Su_15} For measurements at both beamlines, the single-crystal sample was mounted on a thin aluminum plate with tiny amount of GE varnish and put inside a standard closed-cycle cryostat. At HEiDi, a Ge (3 1 1) monochromator was chosen to produce a monochromatic neutron beam with the wavelength of 1.17 Å, and an Er filter was used to minimize the $\lambda$/2 contamination. At DNS, the wavelength of the incident neutrons is 4.2 Å. The {[}0, 1, 0{]} direction of the crystal was aligned perpendicular to the horizontal scattering plane, so that the ($\mathit{H}$, 0, $\mathit{L}$) reciprocal plane can be mapped out by rotating the sample. Throughout this paper, the orthorhombic notation (space group $\mathit{Fmmm}$) will be used for convenience. Single crystals from the same batches were characterized by macroscopic measurements including the resistivity, heat capacity, and dc magnetic susceptibility, using a Quantum Design physical property measurement system (PPMS) and Quantum Design magnetic property measurement system (MPMS).

\section{Experimental Results }

Macroscopic properties of Eu$_{0.5}$Ca$_{0.5}$Fe$_{2}$As$_{2}$ single crystal was shown in Fig. S1 and S2 in the Supplementary Materials. Two magnetic transitions corresponding to the SDW ordering of the Fe sublattice and AFM ordering of the Eu$^{2+}$ moments are identified around 190 K and 10 K, respectively.

To clarify the ground-state magnetic structure of Eu$_{0.5}$Ca$_{0.5}$Fe$_{2}$As$_{2}$, polarized neutron diffraction at 3.5 K was firstly performed at DNS. Fig. 1(a) and 1(b) show the reciprocal-space contour maps of the ($\mathit{H}$, 0, $\mathit{L}$) plane, measured with the neutron polarization parallel to the scattering vector $\mathit{Q}$ ($\mathit{x}$ polarization). The magnetic and nuclear scattering were separated into the spin-flip (SF, Fig. 1(a)) and non-spin-flip (NSF, Fig. 1(b)) channels, respectively.\cite{Scharpf_93} The appearance of (0, 0, 1), (0 ,0 ,3) and (-2, 0, 1) reflections in the $\mathit{X}_{SF}$ channel clearly indicates that the Eu$^{2+}$ moments are antiferromagnetically ordered at 3.5 K, with a propagation vector of $\mathit{k}$ = (0, 0, 1), similar to the undoped parent
compound EuFe$_{2}$As$_{2}$.\cite{Xiao_09} Since magnetic neutron scattering is sensitive to the moment component perpendicular to $\mathit{Q}$, the Eu moments cannot be pointing along the c-axis. Due to imperfection of the polarization analysis, the strong nuclear reflections (0, 0, 2) and (0, 0, 4) observed in the $\mathit{X}_{NSF}$ channel also leaked into the $\mathit{X}_{SF}$ channel. No intensity is observed at (-2, 0, 0) within the experimental resolution, excluding the possibility of a canted-AFM structure of Eu with a net ferromagnetic component along the $\mathit{c}$ axis.\cite{Jin_PhaseDiagram} In addition, magnetic reflections at (-1, 0, 1) and (-1, 0, 3) show up in the $\mathit{X}_{SF}$ channel, arising from the SDW ordering of the Fe moments with the propagation vector of $\mathit{k}$ = (1, 0, 1).\cite{Xiao_09}

\begin{figure}
\centering{}\includegraphics[width=1\columnwidth]{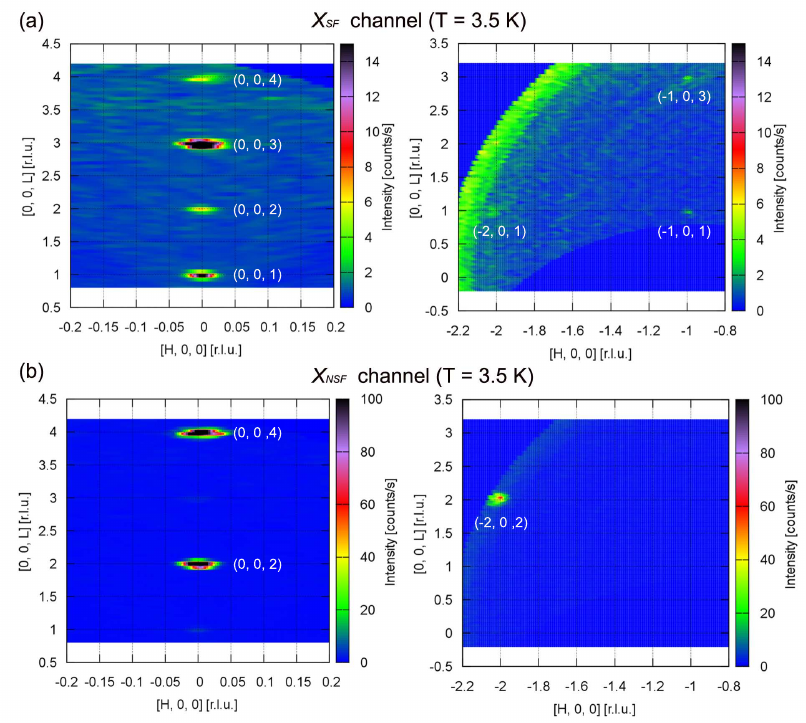}

\caption{Reciprocal-space contour maps in the ($\mathit{H}$, 0, $\mathit{L}$) plane for Eu$_{0.5}$Ca$_{0.5}$Fe$_{2}$As$_{2}$ obtained at T = 3.5 K via polarized neutron diffraction at DNS, with the neutron polarization parallel to the scattering vector $\mathit{Q}$ ($\mathit{x}$ polarization). Intensities in the SF (a) and NSF (b) channels correspond to the magnetic and nuclear reflections, respectively. The appearances of (0, 0, 2) and (0, 0, 4) reflections in the $\mathit{X}_{SF}$ channel are due to the imperfection of the polarization analysis and leakage from the $\mathit{X}_{NSF}$ channel.}
\end{figure}

As shown in the inset of Fig. 2, the (0, 0, 3) magnetic reflection due to the AFM ordering of Eu disappears completely at 11 K in the $\mathit{X}_{SF}$ channel. The temperature dependence of its integrated intensity from the rocking scan is plotted in Fig. 2, indicating an
ordering temperature of $\mathit{T_{Eu}}$= 10.0(5) K, well consistent with that from the macroscopic measurements. The temperature dependence of the integrated intensity, which is proportional to the square of the order parameter, shows a very unusual behaviour. Starting with a negative curvature around $\mathit{T_{Eu}}$, it continues nearly linearly down to the lowest temperature reached in this experiment. Neither typical critical behaviour, nor tendency to saturation can be seen. As discussed below, we attribute this unusual temperature
dependence to the interaction between the Eu and Fe sublattices.

\begin{figure}
\centering{}\includegraphics[width=1\columnwidth]{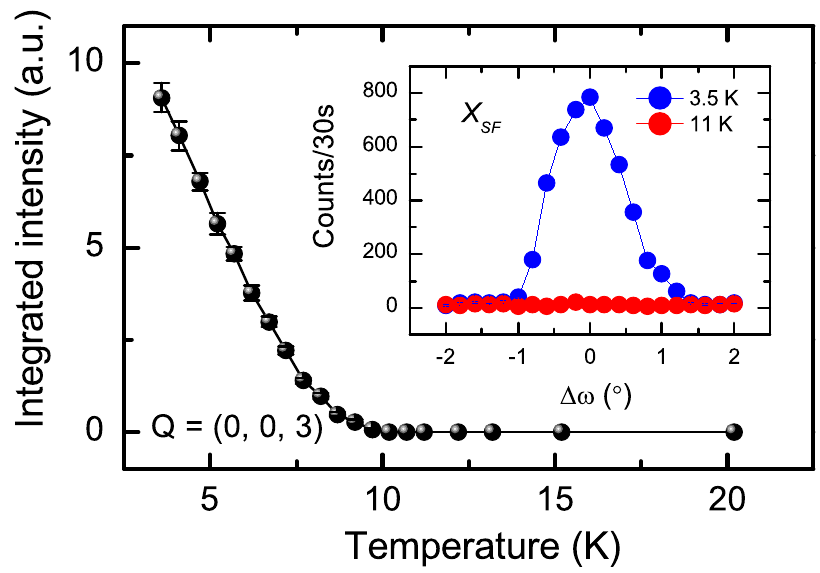}

\caption{The temperature dependence of the integrated intensity of the (0, 0, 3) magnetic reflection in the $\mathit{X}_{SF}$ channel measured at DNS. The insets show the rocking scan of the (0, 0, 3) peak at 3.5 K and 11 K, respectively.}
\end{figure}

The peak intensities of the (1, 0, 1) and (1, 0, 3) reflections in the $\mathit{X}_{SF}$ channel, both arising from the SDW ordering of Fe, were monitored at DNS and plotted using dark spheres in Fig. 3(b) and 3(c), respectively, as a function of the temperature. The onset temperature of the SDW ordering can be estimated to be $\mathit{T_{SDW}}$ = 192(2) K, consistent with that the high-temperature anomaly shown in the resistivity and specific heat data. Interestingly, both order parameters display a kink at the AFM ordering temperature of Eu ($\mathit{T_{Eu}}$ = 10 K). Below $\mathit{T_{Eu}}$, the peak intensity of (1 0 1) reflection increases steadily , while (1 0 3) weakens visibly with decreasing temperature. As shown in the insets of Fig. 3(b) and 3(c), rocking scans of the (1, 0, 1) and (1, 0, 3) reflections in the $\mathit{X}_{SF}$ channel indeed show opposite temperature-dependent tendencies. The temperature dependencies of the integrated intensity of both reflections were also measured at the four-circle neutron diffractometer HEiDi. The same behaviors were observed as shown using the open circles in Fig. 3(b) and 3(c), further confirming the different responses of (1, 0, 1) and (1, 0, 3) reflections to the AFM ordering of the Eu$^{2+}$ moments and suggesting a possible spin-reorientation of the Fe sublattice below $\mathit{T_{Eu}}$. Fig. 3(a) also shows the temperature dependencies of the integrated intensity and full width at half maximum (FWHM) of the (4, 0, 0) nuclear reflection measured at HEiDi. The sudden jump of the intensity and broadening of the peak width indicates the occurrence of a structural phase transition in Eu$_{0.5}$Ca$_{0.5}$Fe$_{2}$As$_{2}$ from a tetragonal (space group $\mathit{I4/mmm}$) to an orthorhombic (space group $\mathit{Fmmm}$) phase at $\mathit{T}_{S}$ = 191(2) K, coincident with the SDW ordering of Fe at $\mathit{T_{SDW}}$, due to the change of extinction conditions of strong nuclear Bragg reflections caused by the emergent orthorhombic domains. 

\begin{figure}
\centering{}\includegraphics[width=1\columnwidth]{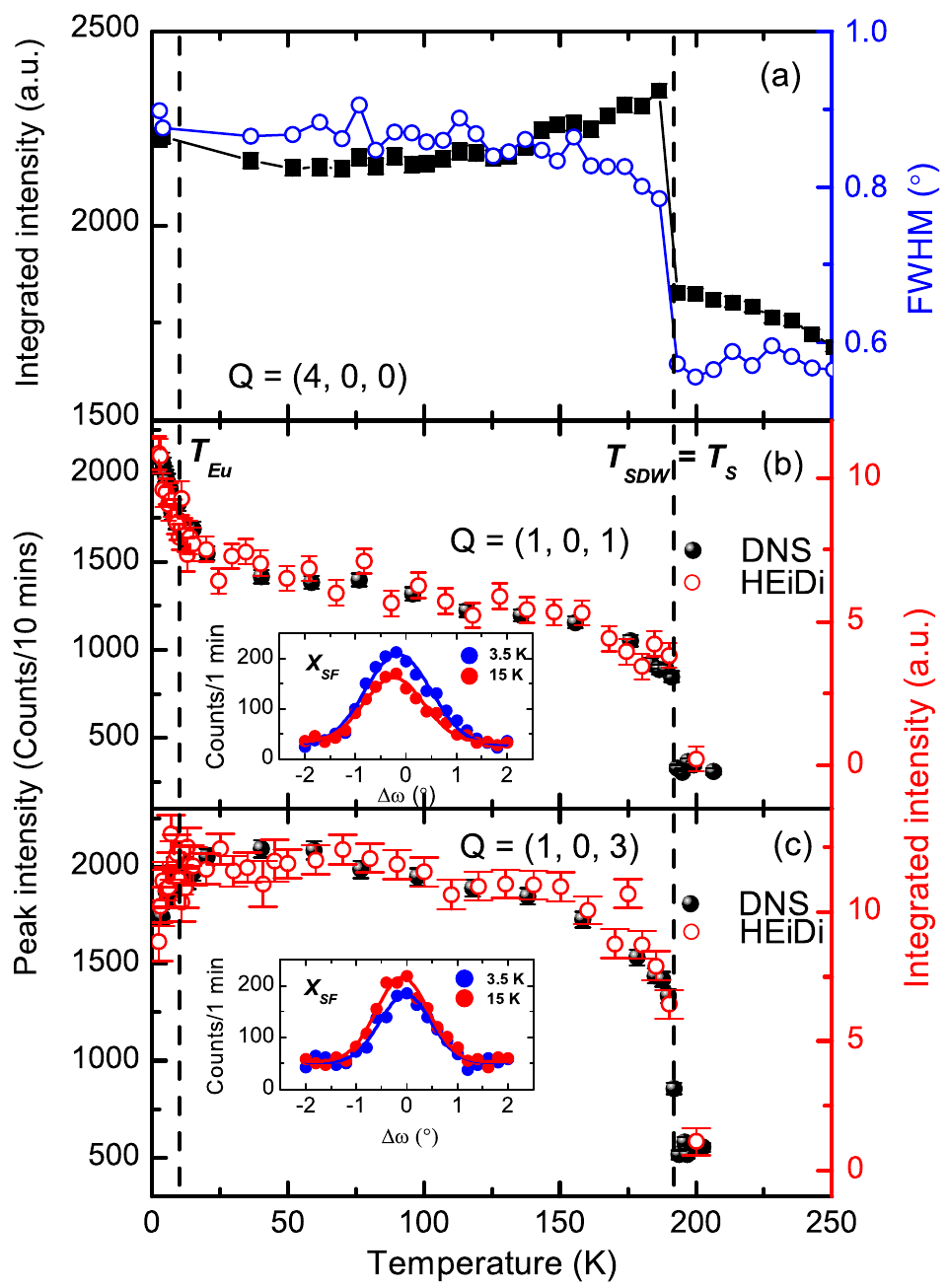}

\caption{The temperature dependencies of the integrated intensity of the (4, 0, 0) nuclear reflection (black squares, a), (1, 0, 1) magnetic reflection (black spheres, b) and (1, 0, 3) magnetic reflection (black spheres, c), respectively, measured at HEiDi. The peak width (FWHM) of (4, 0, 0) is also plotted using blue open circles in (a). Red open circles in (b) and (c) represent the peak intensities of (1, 0, 1) and (1, 0, 3) as a function of the temperature, respectively, measured in the $\mathit{X}_{SF}$ channel at DNS. The insets in (b) and (c) show the rocking scans of (1, 0, 1) and (1, 0, 3) in the $\mathit{X}_{SF}$ channel at 3. 5 K and 15 K, respectively. The solid curves represents the fits using the Gaussian profiles. The dashed vertical lines mark the SDW transition ($\mathit{T_{SDW}}$) coincident with the structural
phase transition ($\mathit{T_{S}}$) and the magnetic ordering of localized Eu$^{2+}$ moments ($\mathit{T}_{Eu}$), respectively.}
\end{figure}

To better understand the interplay of the magnetism between Eu and Fe sublattices in Eu$_{0.5}$Ca$_{0.5}$Fe$_{2}$As$_{2}$, the integrated intensities of 357 nuclear reflections and 254 magnetic reflections from Eu were collected at HEiDi at the base temperature (2.5 K). In
addition, 246 nuclear reflections were collected above $\mathit{T_{Eu}}$ (11 K). The obtained data sets at both temperatures were normalized to the monitor and corrected by the Lorentz factor. After the absorption correction procedure using the DATAP program taking into account the dimensions of the crystal,\cite{Coppens_65} equivalent reflections were merged into the unique ones based on the orthorhombic symmetry. The nuclear structures at 2.5 K and 11 K were refined using FULLPROF.\cite{Rodriguez-Carvajal_93} As shown in Table 1, the nuclear structure does not show a visible difference below and above $\mathit{T_{Eu}}$. At 2.5 K, the magnetic reflections from Eu can be well fitted using the A-type AFM structure as confirmed for the parent compound EuFe$_{2}$As$_{2}$,\cite{Xiao_09} with the Eu$^{2+}$ moment as large as 6.74(4) $\mathit{\mu_{B}}$ aligned along the orthorhombic $\mathit{a}$ axis. Although the Eu sites are diluted with isovalent Ca$^{2+}$ of almost 50 \%, the Eu$^{2+}$ spins are found to be long-range ordered still. This is in good agreement
with a recent $\mu$SR study on the same compound,\cite{Tran_18} and in stark contrast to the short-range magnetic ordering of Eu proposed for hole-doped Eu$_{0.5}$K$_{0.5}$Fe$_{2}$As$_{2}$ based on macroscopic measurements.\cite{Jeevan_08}

\begin{table}
\caption{Refined results for the nuclear structure and Eu magnetic structure of Eu$_{0.5}$Ca$_{0.5}$Fe$_{2}$As$_{2}$ at 2.5 K, as well as the nuclear structure at 11 K. The atomic positions are as follows: Eu/Ca, $4a$ (0, 0, 0); Fe, $8f$ (0.25, 0.25, 0.25); As, $8i$(0, 0, $z$).
The occupancies of Eu and Ca were refined at 2.5 K and fixed at 11 K, to be 48(4)\% and 52(4)\%, respectively. (Space group: $Fmmm$, $\mathit{a}$ = 5.524(2) Å, $\mathit{b}$ = 5.521(1) Å, $\mathit{c}$ = 11.94(1) Å)}

\begin{ruledtabular} %
\begin{tabular}{ccccc}
\multicolumn{2}{c}{} & 2.5 K nuclear & 2.5 K Eu magnetic & 11 K nuclear\tabularnewline
Eu/Ca & $B\,$(Å\textsuperscript{2}) & 0.6(1) &  & 0.6(1)\tabularnewline
 & $M_{a}$($\mu_{B}$) &  & 6.74(4) & \tabularnewline
Fe & $B\,$(Å\textsuperscript{2}) & 0.68(4) &  & 0.66(2)\tabularnewline
As & $z$ & 0.3646(2) &  & 0.3646(2)\tabularnewline
 & $B\,$(Å\textsuperscript{2}) & 0.73(5) &  & 0.70(3)\tabularnewline
\hline 
$R_{F^{2}}$ &  & 6.75 & 14.3 & 6.75\tabularnewline
$R_{wF^{2}}$ &  & 6.14 & 8.49 & 4.51\tabularnewline
$R_{F}$ &  & 3.60 & 10.9 & 3.57\tabularnewline
$\chi^{2}$ &  & 4.89 & 2.31 & 9.43\tabularnewline
\end{tabular}\end{ruledtabular}
\end{table}

Furthermore, motivated by the intriguing responses of the magnetic order parameters of Fe in Eu$_{0.5}$Ca$_{0.5}$Fe$_{2}$As$_{2}$ at $\mathit{T_{Eu}}$, the integrated intensities of a few strong magnetic reflections from the Fe sublattice were collected at HEiDi by performing rocking scans, corrected by the Lorentz factor as well as the absorption effect. Fig.4 shows the integrated intensities of four magnetic reflections with relatively small statistical errors, i.e., (1, 0, 1), (1, 0, 3), (1, 2, 1) and (1, 0, 7). The intensities of them at 11 K can be very well fitted with an in-plane AFM structure of the Fe$^{2+}$ moment (see Fig. 5(a)), with the moment size of $\mathit{M_{a}}$ = 1.10(5) $\mu_{B}$ along the orthorhombic $\mathit{a}$ axis as calculated using FULLPROF. Both the moment direction and moment
size here are quite similar to those observed for the parent compound EuFe$_{2}$As$_{2}$ in the ground state.\cite{Xiao_09} However, the intensities at 2.5 K clearly deviates from those predicted by the in-plane AFM structure. As neutron diffraction only probes the
magnetic moment perpendicular to the scattering vector $\mathit{Q}$, the redistribution of the magnetic scattering intensities signifies a spin reorientation of the Fe$^{2+}$ moments.\cite{Wasser_15} With a canted AFM structure which allows the Fe$^{2+}$ moments to rotate
in the $\mathit{ac}$ plane (see Fig. 5(b)), the intensities at 2.5 K can be well explained with the moment size $\mathit{M_{a}}$ = 0.85(5) $\mu_{B}$ and $\mathit{M_{c}}$ = 0.22(5) $\mu_{B}$. All the details about the model refinements of the magnetic structure of Fe were included in the Supplementary Materials.

\begin{figure}
\centering{}\includegraphics[width=1\columnwidth]{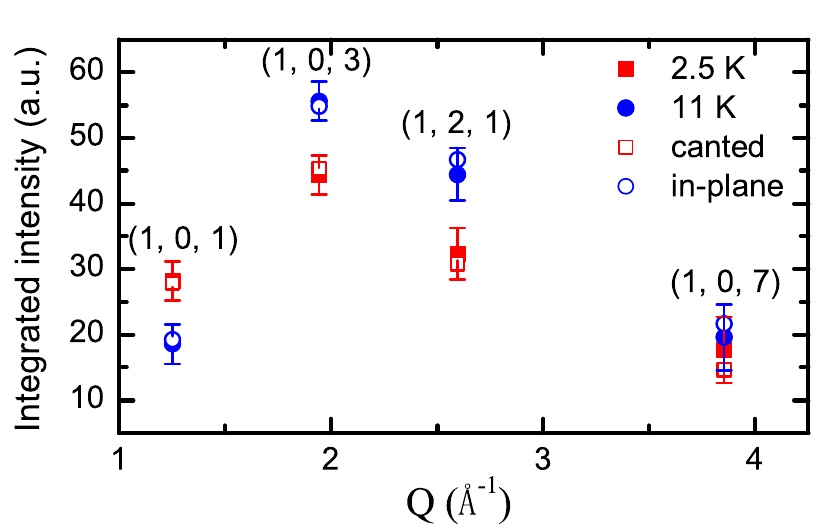}

\caption{Comparison between the observed intensities of four magnetic reflections from the Fe sublattice at 11 K (blue filled circles), 2.5 K (red filled squares), the calculated intensities using the in-plane AFM structure (blue open circles), and the calculated intensities using the canted AFM structure (red open squares).}
\end{figure}

The magnetic structures of Eu$_{0.5}$Ca$_{0.5}$Fe$_{2}$As$_{2}$ at 11 K and 2.5 K are illustrated in Fig. 5(a) and 5(b), respectively. As concluded above, at 11 K (slightly above $\mathit{T_{Eu}}$), the Eu sublattice is not magnetically ordered yet while the Fe sublattice
displays an in-plane AFM structure. With further decreasing temperature, the Eu$^{2+}$ moments start to order, while the Fe$^{2+}$ moments tend to rotate towards the $\mathit{c}$ axis within the $\mathit{ac}$ plane, as reflected by the opposite temperature-dependent tendencies of its magnetic order parameters shown in Fig. 3. At the reachable base temperature (2,5 K) at HEiDi, the Eu$^{2+}$ spins are found to align along the $\mathit{a}$ axis in the A-type AFM structure similar to the undoped parent compound, while the Fe$^{2+}$ moments are ordered in a canted AFM structure with a canting angle of 14(4)° out of the $\mathit{ab}$ plane. In other words, the spin reorientation of the Fe$^{2+}$ moments occurs in coincidence with the AFM ordering of Eu, suggesting a strong interplay between the two magnetic sublattices in Eu$_{0.5}$Ca$_{0.5}$Fe$_{2}$As$_{2}$.

\begin{figure}
\centering{}\includegraphics[width=1\columnwidth]{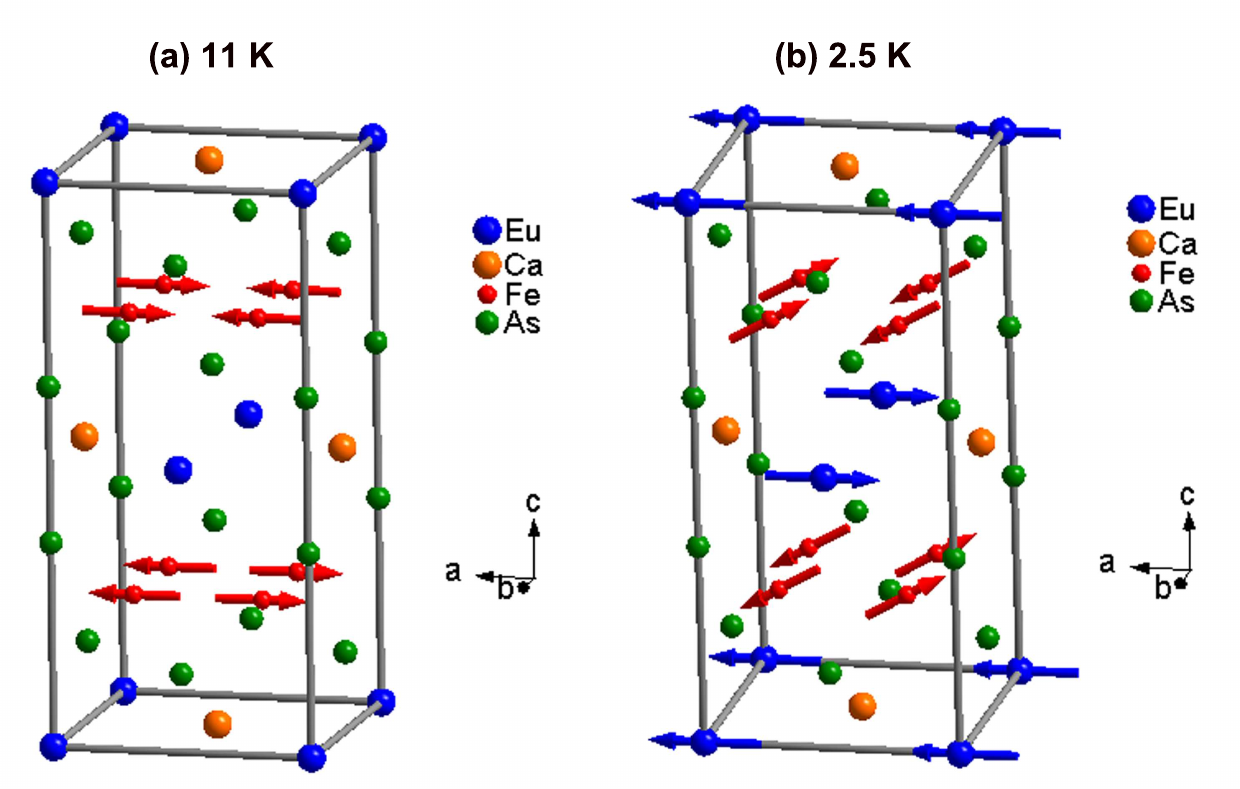}

\caption{The magnetic structures of Eu$_{0.5}$Ca$_{0.5}$Fe$_{2}$As$_{2}$ at 11 K (a) and 2.5 K (b), respectively.}
\end{figure}

\section{Discussion and Conclusion}

We note that the magnetic order parameter of the Eu sublattice in Fig. 2 does not saturate down to 3.5 K. However, refinements using the magnetic reflections from Eu yield the moment size of 6.74(4) $\mu_{B}$, well consistent with a full moment of $\mu_{S}$ = $\mathit{gS}$
= 7 $\mu_{B}$ expected for the Eu$^{2+}$ spins with $\mathit{S}$ = 7/2. This means that the Eu sublattice is fully magnetically ordered at 3.5 K. Thus, the unusual temperature dependence in Fig. 2 is very likely to arise from the change of the strength of Eu-Fe magnetic
interaction concomitant with the spin reorientation of the Fe$^{2+}$ spins moments. As well documented, the coupling between two AFM sublattices may arise from quantum fluctuations via the so-called ``order-by-disorder'' mechanism.\cite{Brueckel_88,Brueckel_92,Brueckel_95,Gukasov_88} The strong Eu-Fe coupling in Eu$_{0.5}$Ca$_{0.5}$Fe$_{2}$As$_{2}$, therefore, might be due to the longitudinal fluctuations of the Eu$^{2+}$ spins, which lead to a considerable change in the magnetic anisotropy
energy and result in the spin reorientation of the Fe$^{2+}$ moments. For the EuFe$_{2}$As$_{2}$ system, to the best of our knowledge, our observation here provides the first experimental evidence of spin reorientation of Fe below the Eu magnetic ordering temperature. Eu$_{0.5}$Ca$_{0.5}$Fe$_{2}$As$_{2}$, therefore, exhibits a strong Eu-Fe interplay undoubtly. However, this is not contradictory with the weak Eu-Fe coupling in the parent compound EuFe$_{2}$As$_{2}$,\cite{Xiao_09,Herrero-Martin_09,Jeevan_08_parent}
since it was found that the strength of interplay between 3$\mathit{d}$ and 4$\mathit{f}$ electrons can be tunable by chemical doping.\cite{Shang_13} Compared with undoped EuFe$_{2}$As$_{2}$, the out-of-plane lattice constant $\mathit{c}$ shrinks considerably by \textasciitilde{} 1 \% upon 50 \% Ca substitution. As a result, the neareast Eu-Fe distance reduces by 0.7\% from 3.591 Å (in EuFe$_{2}$As$_{2}$) to 3.567 Å (in Eu$_{0.5}$Ca$_{0.5}$Fe$_{2}$As$_{2}$), favoring a stronger Eu-Fe spin interaction. Further theoretical studies on Eu$_{0.5}$Ca$_{0.5}$Fe$_{2}$As$_{2}$ will be very helpful for understanding its intriguing magnetic properties and strong Eu-Fe interplay in it.

In conclusion, using complementary polarized and unpolarized single-crystal neutron diffraction, we have investigated the temperature-dependent magnetic structures of Eu$_{0.5}$Ca$_{0.5}$Fe$_{2}$As$_{2}$. Upon 50 \% dilution of the Eu sites with isovalent Ca$^{2+}$, the Eu sublattice is found to be still long-range ordered below $\mathit{T_{Eu}}$ = 10 K, in the A-typed AFM structure. The moment size of Eu$^{2+}$ spins is estimated to be as large as 6.74(4) $\mu_{B}$ at 2.5 K. The Fe sublattice undergoes a SDW transition at $\mathit{T_{SDW}}$ = 192(2) K and displays an in-plane AFM structure above $\mathit{T_{Eu}}$. However, at 2.5 K, the Fe$^{2+}$ moments are found to be ordered in a canted AFM structure with a canting angle of 14(4)° out of the $\mathit{ab}$ plane. The spin reorientation of Fe below the AFM ordering temperature of Eu provides a direct evidence of a strong interplay between the two magnetic sublattices in Eu$_{0.5}$Ca$_{0.5}$Fe$_{2}$As$_{2}$.

\bibliographystyle{apsrev} \bibliographystyle{apsrev}
\begin{acknowledgments}
W. T. J. would like to acknowledge S. Mayr for the assistance with the orientation of the crystal. This work is based on experiments performed at the HEiDi and DNS instrument operated by Jülich Centre for Neutron Science (JCNS) at the Heinz Maier-Leibnitz Zentrum (MLZ),
Garching, Germany. Z. B. acknowledges financial support from National Science Center, Poland Grant 2017/25/B/ST3/02868.
\bibliographystyle{apsrev}

\begin{thebibliography}{10}

\bibitem{Kamihara_08}
Y. Kamihara, T. Watanabe, M. Hirano, and H. Hosono, J. Am. Chem. Soc. \textbf{130}, 3296 (2008).

\bibitem{ChenXH_08}
X. H. Chen, T. Wu, G. Wu, R. H. Liu, H. Chen, and D. F. Fang, Nature \textbf{453}, 761 (2008).

\bibitem{ChenGF_08}
G. F. Chen, Z. Li, D.Wu, G. Li,W. Z. Hu, J. Dong, P. Zheng, J. L. Luo, and N. L. Wang, Phys. Rev. Lett. \textbf{100}, 247002 (2008).

\bibitem{RenZA_08}
Z.-A. Ren, G.-C. Che, X.-L. Dong, J. Yang, W. Lu, W. Yi, X. -L. Shen, Z.-C. Li, L.-L. Sun, F. Zhou, et al., Europhys. Lett. \textbf{83}, 17002 (2008).

\bibitem{WangC_08}
C. Wang, L. Li, S. Chi, Z. Zhu, Z. Ren, Y. Li, Y. Wang, X. Lin, Y. Luo, S. Jiang, et al., Europhys. Lett. \textbf{83}, 67006 (2008).

\bibitem{Rotter_08}
M. Rotter, M. Tegel, and D. Johrendt, Phys. Rev. Lett. \textbf{101}, 107006 (2008).

\bibitem{Sefat_08}
A. S. Sefat, R. Jin, M. A. McGuire, B. C. Sales, D. J. Singh, and D. Mandrus, Phys. Rev. Lett. \textbf{101}, 117004 (2008).

\bibitem{Zapf_17}
S. Zapf and M. Dressel, Rep. Prog. Phys. \textbf{80}, 016501 (2017).

\bibitem{Johnston_10}
D. C. Johnston, Adv. Phys. \textbf{59}, 803 (2010).

\bibitem{Dai_15}
P. Dai, Rev. Mod. Phys. \textbf{87}, 855 (2015).

\bibitem{Ryan_09}
D. H. Ryan, J. M. Cadogan, C. Ritter, F. Canepa, A. Palenzona, and M. Putti, Phys. Rev. B \textbf{80}, 220503 (2009).

\bibitem{Ren_09}
Z. Ren, Q. Tao, S. Jiang, C. Feng, C. Wang, J. Dai, G. Cao, and Z. Xu, Phys. Rev. Lett. \textbf{102}, 137002 (2009).	

\bibitem{Jeevan_11}
H. S. Jeevan, D. Kasinathan, H. Rosner, and P. Gegenwart, Phys. Rev. B \textbf{83}, 054511 (2011).	

\bibitem{Guguchia_13}
Z. Guguchia, A. Shengelaya, A. Maisuradze, L. Howard, Z. Bukowski, M. Chikovani, H. Luetkens, S. Katrych, J. Karpinski, and H. Keller, J. Supercond. Nov. Magn. \textbf{26}, 285 (2013).

\bibitem{Maeter_09}
H. Maeter, H. Luetkens, Y. G. Pashkevich, A. Kwadrin, R. Khasanov, A. Amato, A. A. Gusev, K. V. Lamonova, D. A. Chervinskii, R. Klingeler, et al., Phys. Rev. B \textbf{80}, 094524 (2009).

\bibitem{ZhangQ_13}
Q. Zhang, W. Tian, H. Li, J.-W. Kim, J. Yan, R. W. McCallum, T. A. Lograsso, J. L. Zarestky, S. L. Bud’ko, R. J. McQueeney, et al., Phys. Rev. B \textbf{88}, 174517 (2013).

\bibitem{Nandi_11}
S. Nandi, Y. Su, Y. Xiao, S. Price, X. F. Wang, X. H. Chen, J. Herrero-Martín, C. Mazzoli, H. C. Walker, L. Paolasini, et al., Phys. Rev. B \textbf{84}, 054419 (2011).

\bibitem{Stockert_12}
U. Stockert, N. Leps, L.Wang, G. Behr, S.Wurmehl, B. Büchner, and R. Klingeler, Phys. Rev. B \textbf{86}, 144407 (2012).

\bibitem{Tian_10}
W. Tian, W. Ratcliff, M. G. Kim, J.-Q. Yan, P. A. Kienzle, Q. Huang, B. Jensen, K. W. Dennis, R. W. McCallum, T. A. Lograsso, et al., Phys. Rev. B \textbf{82}, 060514 (2010).

\bibitem{Xiao_09}
Y. Xiao, Y. Su, M. Meven, R. Mittal, C. M. N. Kumar, T. Chatterji, S. Price, J. Persson, N. Kumar, S. K. Dhar, A. Thamizhavel, and Th. Brueckel, Phys. Rev. B \textbf{80}, 174424 (2009).	

\bibitem{Herrero-Martin_09}
J. Herrero-Mart\'in, V. Scagnoli, C. Mazzoli, Y. Su, R. Mittal, Y. Xiao, Th. Brueckel, N. Kumar, S. K. Dhar, A. Thamizhavel, and L. Paolasini, Phys. Rev. B \textbf{80}, 134411 (2009).	

\bibitem{Jeevan_08_parent}
H. S. Jeevan, Z. Hossain, D. Kasinathan, H. Rosner, C. Geibel, and P. Gegenwart, Phys. Rev. B \textbf{78}, 052502 (2008).

\bibitem{Pogrebna_15}
A. Pogrebna, T. Mertelj, N. Vujiˇci´c, G. Cao, Z. A. Xu, and D. Mihailovic, Sci. Rep. \textbf{5}, 7754 (2015)

\bibitem{Ahmed_10}
A. Ahmed, M. Itou, S. Xu, Z. Xu, G. Cao, Y. Sakurai, J. Penner-Hahn, and A. Deb, Phys. Rev. Lett. \textbf{105}, 207003 (2010).

\bibitem{Guguchia_11}
Z. Guguchia, J. Roos, A. Shengelaya, S. Katrych, Z. Bukowski, S.Weyeneth, F. Murányi, S. Strässle, A. Maisuradze, J. Karpinski, et al., Phys. Rev. B \textbf{83}, 144516 (2011).

\bibitem{Blachowski_11}
A. B\l{}achowski, K. Ruebenbauer, J. \ifmmode \dot{Z}\else \.{Z}\fi{}ukrowski, Z. Bukowski, K. Rogacki, P. J. W. Moll, and J. Karpinski, Phys. Rev. B \textbf{84}, 174503 (2011).	

\bibitem{Jin_16}
W. T. Jin, Y. Xiao, Y. Su, S. Nandi, W. H. Jiao, G. Nisbet, S. Demirdis, G. H. Cao, and T. Br\"uckel, Phys. Rev. B \textbf{93}, 024517 (2016).	

\bibitem{Xiao_10}
Y. Xiao, Y. Su, W. Schmidt, K. Schmalzl, C. M. N. Kumar, S. Price, T. Chatterji, R. Mittal, L. J. Chang, S. Nandi, et al., Phys. Rev. B (R) \textbf{81}, 220406 (2010).

\bibitem{Xiao_12}
Y. Xiao, Y. Su, S. Nandi, S. Price, B. Schmitz, C. M. N. Kumar, R. Mittal, T. Chatterji, N. Kumar, S. K. Dhar, et al., Phys. Rev. B \textbf{85}, 094504 (2012).

\bibitem{Jin_13}
W. T. Jin, S. Nandi, Y. Xiao, Y. Su, O. Zaharko, Z. Guguchia, Z. Bukowski, S. Price, W. H. Jiao, G. H. Cao, and Th. Br\"uckel, Phys. Rev. B \textbf{88}, 214516 (2013).	

\bibitem{Nandi_14}
S. Nandi, W. T. Jin, Y. Xiao, Y. Su, S. Price, D. K. Shukla, J. Strempfer, H. S. Jeevan, P. Gegenwart, and Th. Br\"uckel, Phys. Rev. B \textbf{89}, 014512 (2014).	

\bibitem{Nandi_14_neutron}
S. Nandi, W. T. Jin, Y. Xiao, Y. Su, S. Price, W. Schmidt, K. Schmalzl, T. Chatterji, H. S. Jeevan, P. Gegenwart, and Th. Br\"uckel, Phys. Rev. B \textbf{90}, 094407 (2014).	

\bibitem{Jin_15}
W. T. Jin, W. Li, Y. Su, S. Nandi, Y. Xiao, W. H. Jiao, M. Meven, A. P. Sazonov, E. Feng, Y. Chen, C. S. Ting, G. H. Cao, and Th. Br\"uckel, Phys. Rev. B \textbf{91}, 064506 (2015).

\bibitem{Jin_PhaseDiagram}
W. T. Jin, Y. Xiao, Z. Bukowski, Y. Su, S. Nandi, A. P. Sazonov, M. Meven, O. Zaharko, S. Demirdis, K. Nemkovski, et al., Phys. Rev. B \textbf{94}, 184513 (2016).

\bibitem{Jin_Pressure}
W. T. Jin, J. P. Sun, G. Z. Ye, Y. Xiao, Y. Su, K. Schmazl, S. Nandi, Z. Bukowski, Z. Guguchia, E. Feng, et al., Sci. Rep. \textbf{7}, 3532 (2017).

\bibitem{Marchand_78}
R. Marchand and W. Jeitschko, J. Solid State Chem. \textbf{24}, 351 (1978).	

\bibitem{Jiang_09_NJP}
S. Jiang, H. Xing, G. Xuan, Z. Ren, C. Wang, Z. A. Xu, and G. Cao, Phys. Rev. B \textbf{80}, 184514 (2009).	

\bibitem{Mitsuda_11}
A. Mitsuda, S. Seike, T. Matoba, H.Wada, F. Ishikawa, and Y. Yamada, J. Phys. Conf. Ser. \textbf{273}, 012100 (2011).

\bibitem{Tran_18}
L. M. Tran, M. Babij, L. Korosec, T. Shang, Z. Bukowski, and T. Shiroka, Phys. Rev. B \textbf{98}, 104412 (2018).

\bibitem{Harnagea_18}
L. Harnagea, R. Kumar, S. Singh, S. Wurmehl, A. U. B. Wolter, and B. Büchner, J. Phys. Condens. Matter \textbf{30}, 415601 (2018).

\bibitem{Meven_15}
M. Meven and A. Sazonov, J. Large-Scale Res. Facilities \textbf{1}, A7 (2015), URL http://dx.doi.org/10.17815/jlsrf-1-20.

\bibitem{Su_15}
Y. Su, K. Nemkovskiy, and S. Demirdis, J. Large-Scale Res. Facilities \textbf{1}, A27 (2015), URL http://dx.doi.org/10.17815/jlsrf-1-33.

\bibitem{Jeevan_08}
H. S. Jeevan, Z. Hossain, D. Kasinathan, H. Rosner, C. Geibel, and P. Gegenwart, Phys. Rev. B \textbf{78}, 092406 (2008).

\bibitem{Jin_EuNi}
W. T. Jin, N. Qureshi, Z. Bukowski, Y. Xiao, S. Nandi, M. Babij, Z. Fu, Y. Su, and T. Brückel, Phys. Rev. B \textbf{99}, 014425 (2019).

\bibitem{Komedera_18}
K. Komedera, A. Błachowski, K. Ruebenbauer, J. Zukrowski, S. M. Dubiel, L. M. Tran, B. M., and Z. Bukowski, J. Magn. Magn. Mater. \textbf{457}, 1 (2018).

\bibitem{Scharpf_93}
O. Schärpf and H. Capellmann, Phys. Stat. Sol. A \textbf{135}, 359 (1993).

\bibitem{Coppens_65}
P. Coppens, L. Leiserowitz, and D. Rabinovich, Acta Crystallogr. \textbf{18}, 1035 (1965).

\bibitem{Rodriguez-Carvajal_93}
J. Rodr\'iguez-Carvajal, Physica B \textbf{192}, 55 (1993).	

\bibitem{Wasser_15}
F. Waßer, A. Schneidewind, Y. Sidis, S. Wurmehl, S. Aswartham, B. Büchner, and M. Braden, Phys. Rev. B \textbf{91}, 060505 (2015).

\bibitem{Brueckel_88}
T. Brueckel, B. Dorner, A. G. Gukasov, V. P. Plakhty, W. Prandl, E. F. Shender, and O. P. Smirnow, Z. Phys. B \textbf{72}, 477 (1988).

\bibitem{Brueckel_92}
T. Brückel, B. Dorner, A. Gukasov, and V. Plakhty, Phys. Lett. A \textbf{162}, 357 (1992).

\bibitem{Brueckel_95}
T. Brueckel, C. Paulsen, K. Hinrichs, and W. Prandl, Z. Phys. B \textbf{97}, 391 (1995).

\bibitem{Gukasov_88}
A. G. Gukasov, T. Brückel, B. Dorner, V. P. Plakhty, W. Prandl, E. F. Shender, and O. P. Smirnov, Europhys. Lett. \textbf{7}, 83 (1988).

\bibitem{Shang_13}
T. Shang, L. Yang, Y. Chen, N. Cornell, F. Ronning, J. L. Zhang, L. Jiao, Y. H. Chen, J. Chen, A. Howard, et al., Phys. Rev. B \textbf{87}, 075148 (2013).

\end{thebibliography}
\end{acknowledgments}

\end{document}